\documentclass[10pt,two column]{article}
\usepackage[utf8]{inputenc}
\usepackage[T1]{fontenc}
\usepackage[hidelinks]{hyperref}
\usepackage{url}
\usepackage{float}
\usepackage{booktabs}
\usepackage{amsfonts}
\usepackage{nicefrac}
\usepackage{microtype}
\usepackage{graphicx}
\usepackage{amsmath}
\usepackage{tikz}
\usepackage{import}
\usepackage{lipsum}

\newcommand{\norm}[1]{\left\lVert#1\right\rVert}
\newcommand*\samethanks[1][\value{footnote}]{\footnotemark[#1]}
\title{Learning protein conformational space by enforcing \\ physics with convolutions and latent interpolations}

\author{Venkata K. Ramaswamy$^1$\thanks{The authors contributed equally.} \and Chris G. Willcocks$^2$\samethanks \and Matteo T. Degiacomi$^1$\samethanks}
\date{
    $^1$Department of Physics, Durham University\\
    $^2$Department of Computer Science, Durham University
}

\begin{document}
\maketitle
\begin{abstract}
Determining the different conformational states of a protein and the transition paths between them is key to fully understanding the relationship between biomolecular structure and function. This can be accomplished by sampling protein conformational space with molecular simulation methodologies. Despite advances in computing hardware and sampling techniques, simulations always yield a discretized representation of this space, with transition states undersampled proportionally to their associated energy barrier. We present a convolutional neural network that learns a continuous conformational space representation from example structures, and loss functions that ensure intermediates between examples are physically plausible. We show that this network, trained with simulations of distinct protein states, can correctly predict a biologically relevant non-linear transition path, without any example on the path provided. We also show we can transfer features learnt from one protein to others, which results in superior performances, and requires a surprisingly small number of training examples. 

\end{abstract}

\section{Introduction}

The shape of a protein determines its capacity of interacting, under given environmental conditions, with specific binding partners as different as ions, DNA, lipids, drugs or other proteins. These tightly controlled interactions are key for life as we know it. Proteins should however not be considered as a single atomic arrangement, as their relationship with the environment (e.g. temperature, pressure, pH, binding to other molecules) results in specific conformational dynamics. These dynamics define a continuous conformational space, often conveniently subdivided into a discrete set of low energy states and higher energy transition paths between them. A full understanding of the function of any protein in an organism thus requires accurate knowledge of its conformational space. Proteins are often composed of thousands of atoms, that should in principle be associated with an enormous amount of possible arrangements. While only a small energetically favourable fraction of these arrangements is accessible~\cite{zwanzig1992levinthal}, only a subset of these, those of lowest energy, can be typically observed at atomic resolution by experimental techniques. In order to obtain a larger amount of conformational space samples, simulation techniques such as molecular dynamics (MD) can be exploited. These techniques iteratively generate new structures based on an initial, known atomic arrangement and a physical model of atomic interactions. However, biologically relevant conformational changes can sometimes take place in the timescale of milliseconds or higher (e.g. the full cycle of GroEL chaperonin from its closed to open state takes $\sim$15s~\cite{karplus2002molecular}) which are beyond the reach of conventional MD.

Deep neural networks are able to learn continuous representations that capture the structure of a dataset. In particular, generative models such as variational autoencoders~\cite{kingma2013autoencoding}, generative adversarial networks~\cite{goodfellow2014gan} or Boltzmann generators~\cite{noe2019boltzmann} have been showing a remarkable ability to synthesize complex and sparse datasets. Generative neural networks create an internal model recapitulating example data, a model that can then be interrogated to generate new, plausible data samples. While most generative models produce new samples from an assumed prior distribution, recent architectures improve interpolations through additional adversarial components~\cite{berthelot2018understanding}. Many successful generative architectures utilise convolutional neural networks (CNNs)~\cite{lecun2015deep}. These are more computationally efficient than regular neural networks and have fewer parameters due to shared weights mimicking local connectivity in the visual cortex of the brain. Though a surfeit of applications of CNNs are in the fields of image, video, audio, and speech recognition~\cite{lecun2015deep}, variants have also been applied to bioinformatics ranging from gene expression regulation~\cite{alipanahi2015predicting, zhou2015predicting, denas2013deep}, anomaly classification~\cite{Roth2016improving, ypsilantis2015predicting, zeng2015deep}, to prediction of protein secondary structure~\cite{zhang2018prediction, long2019protein}, and protein folds~\cite{Hou2017DeepSF, zamora2019structural, gao2019destini, evans2018novo}. 

\begin{figure*}[ht!]
  \includegraphics[width=\textwidth]{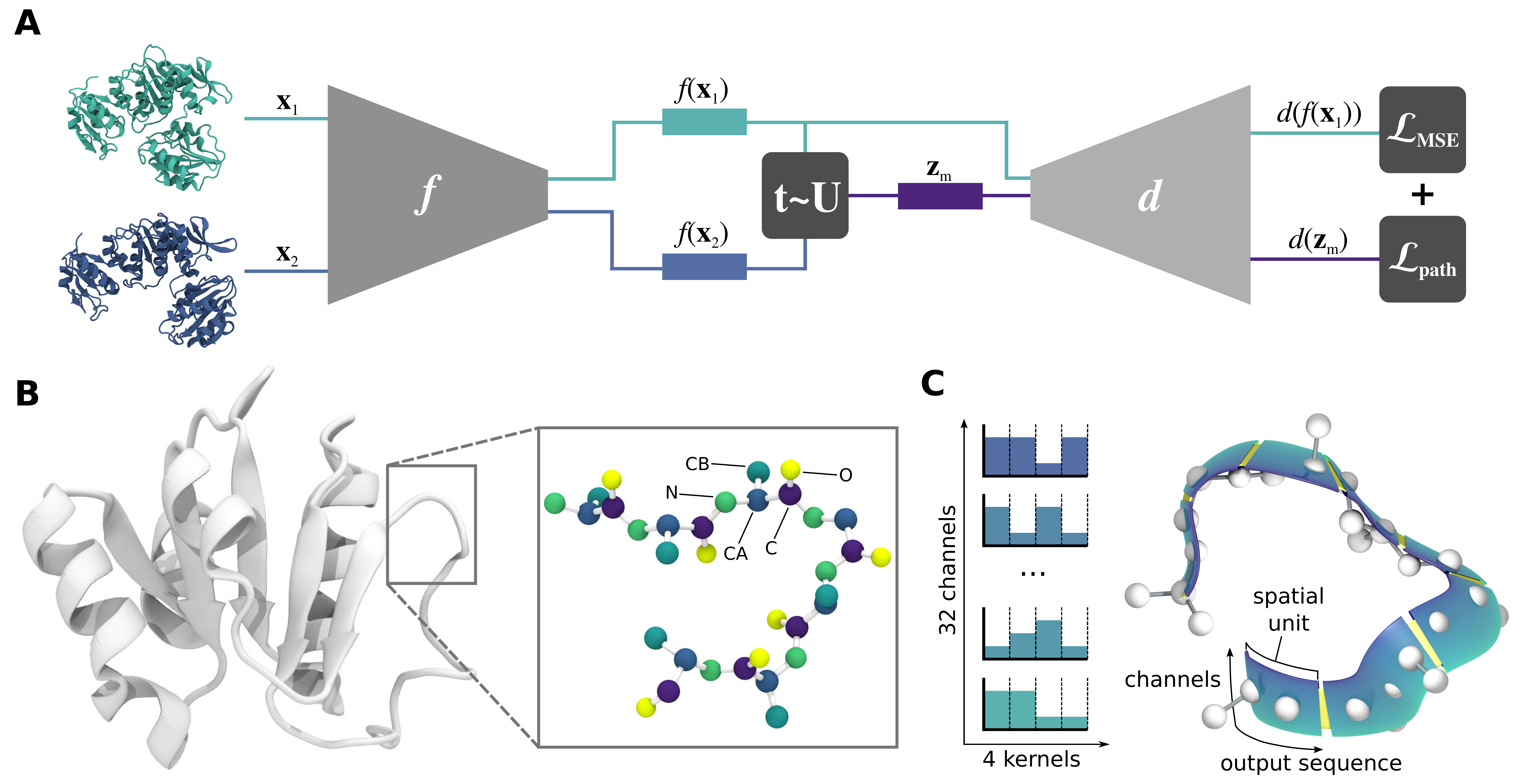}
  \vspace{-1em}\caption{Neural network design. (A) The generative architecture is composed of an encoder $f$ and a decoder $d$, and is trained with a collection of protein conformations. The loss function couples a geometric term $\mathcal{L}_\text{MSE}$ to ensure original and encoded-decoded structure are similar, and convolutional terms $\mathcal{L}_\text{path}$ to ensure that latent space interpolations $z_m$ between any pair of conformations produce protein structures of low energy. (B) Protein atoms can be sorted in a list so that atoms that are adjacent in the list are also adjacent in the Cartesian space. The convolutional neural network operates on this list, that can be of arbitrary size. (C) The first 1D convolution layer learns $32 \times$ feature detectors, each with a kernel size of $4$. The stride is set to $2$, therefore the output sequence is half the input size for any input length. Each subsequent layer further reduces the spatial length of the molecule, warping the input such that it becomes progressively deeper and thicker (more ribbon-like) as well as more abstract.}
  \label{fig:conv}
\end{figure*}

We present a 1D CNN architecture (Figure \ref{fig:conv}A), directly trainable with protein structures to build a model of their underlying conformational space. To improve the generalisation capability of our network, we design a new loss function that leverages on, as a prior, knowledge of physical laws dictating atomic interactions. Our architecture and training approach leads to several significant advantages over conventional networks taking atomic coordinates~\cite{degiacomi2019coupling} or molecular features as input. First, being fully convolutional, our architecture features a small number of parameters and is therefore easy to train. Second, it can handle input molecules with arbitrary numbers of atoms, enabling network training with different molecules, either simultaneously or via transfer learning. Third, it does not make assumptions on the distribution of data around observations used for training. We show that these features enable us to identify biologically relevant intermediate protein configurations along plausible transition paths between known low-energy states.

\section{Results}

MurD (UDP-N-acetylmuramoyl-L-alanine:D-glutamate ligase)~\cite{bertrand1997crystal} plays a key role in the peptidoglycan biosynthesis of almost all bacterial species by catalyzing the addition of D-glutamic acid to UDP-N-acetylmuramoyl-L-alanine. The protein consists of three globular domains, one of which undergoes a large scale rearrangement (from open to closed state) triggered by substrate binding to activate the catalytic site. The open (PDB: 1E0D~\cite{bertrand2000open}) and closed (PDB: 3UAG~\cite{bertrand1999determination}) states, as well as a few intermediates (PDB: 5A5E and 5A5F~\cite{vsink2016crystallographic}), have been crystallized, providing key experimental evidence about the possible protein's mode of action. MD simulations of the open and closed states, and of the transition between them have been previously carried out~\cite{degiacomi2019coupling}, providing a useful dataset for our network training and its performance evaluation. Such extensive data and the importance of MurD as a potential antibacterial drug target~\cite{vsink2013murd, belete2019novel} make this protein a particularly interesting test case for this study. 

Here, we train our neural network with conformations of MurD open and closed states generated by MD (\textit{training set}) and assess the network capacity of predicting a possible transition path between the two. We assess the quality of the predicted path in terms of its structural quality as well as matching with the closed-to-open MD simulation and available intermediate crystal structures. We then evaluate the capacity of our network trained on MurD to adapt to other proteins in a transfer learning scenario.

\subsection{Force field-based loss functions improve network accuracy}

\begin{figure*}[ht!]
    \includegraphics[width=\textwidth]{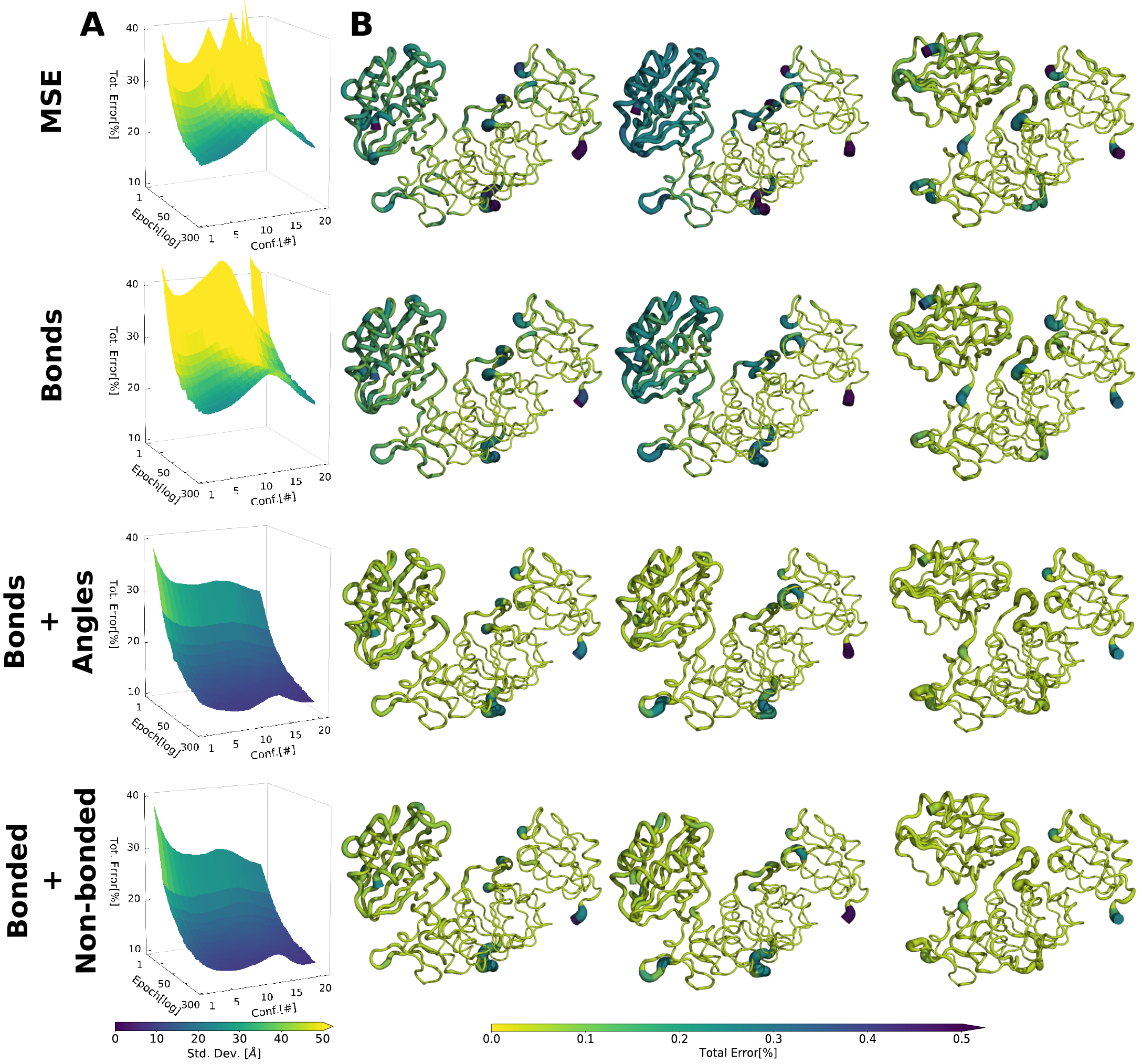}
    \vspace{-1em}\caption{Performance of the neural network trained with four different loss functions featuring an increasing number of physical terms (labelled on the left). At every epoch, each network was asked to generate 20 conformations interpolating from MurD closed to open state. (A) At each epoch, we calculate the total error of each conformation, defined as the sum of the difference between all the measured bond and angles and their accepted (force field-based) values. Mean values are reported on the vertical axes, standard deviations in colour. Physics-based loss functions lead to interpolations with lower errors. (B) The network-predicted protein conformations of open (left), intermediate (centre) and closed (right) states at the last epoch, shown in sausage representation with the thickness and colour corresponding to the percentage error at the residue level.}
    \label{fig:toterr}
\end{figure*}

We first assessed whether training the neural network using information on the physical properties of proteins is beneficial. To this end, we trained it with four different loss functions featuring an increasing amount of physics-based constraints (Figure \ref{fig:toterr}). Each network was trained with conformations produced by the MD simulations of MurD closed and open states and then challenged to predict a transition path by producing 20 intermediates between the two states. To assess the quality of intermediate conformations throughout the training, we measured each of their bonds and angles and compared these measures to the expected equilibrium values within the Amber ff14SB force field~\cite{maier2015ff14sb}. Thus, for each intermediate conformation at every training epoch, we determined a total error in the bonded parameters (see Methods section and Figures S1 and S2).

The network trained without any physics-based constraints (using just the Mean Square Error, MSE) could only poorly reproduce MurD expected bonded parameters (average error of intermediate structures equal to $\sim$22.9\% $\pm$ 2.3). Notably, the interpolation quality degraded the further the intermediate structure was from examples within the training set (errors of up to 26.6\%). Regions of the protein featuring the highest error were concentrated in loops and on the mobile domain of the protein (Figure~\ref{fig:toterr}B: MSE).

Coupling bond information with MSE in the loss function only slightly improved the quality of the protein structures generated (Figure \ref{fig:toterr}: Bonds), with average errors equal to 23.3\% $\pm$ 2.3. A substantial improvement was then obtained when training a network with MSE, bonds and angles. Intermediate conformations featured an average error of 12.9\% $\pm$ 0.8, with a worst-case of 14.4\% (Figure \ref{fig:toterr}: Bonds + Angles). Mapping bond and angle errors on MurD structure revealed that the only regions of lower quality were within a few loops.

\begin{figure*}[ht!]
    \includegraphics[width=\textwidth]{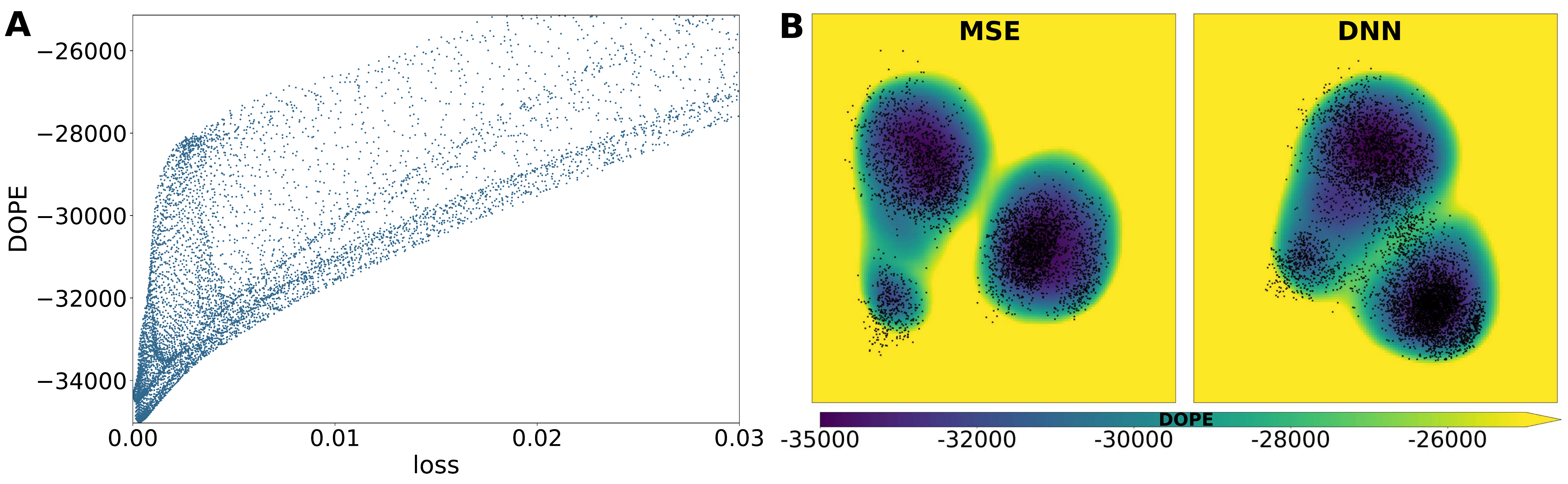}
    \vspace{-1em}\caption{Analysis of latent spaces of networks trained on conformations of MurD "open" and "closed" states. (A) The physics-based loss function used to train our neural network correlates with the DOPE score, making it a reasonable estimator of protein structural quality. (B) DOPE scores corresponding to the latent spaces of our neural network trained with two different loss functions. On the left, the network is trained to only minimise mean square error (MSE) between input and output structure, on the right the loss function combines MSE and a physics-based loss function (DDN). Yellow regions indicate structures of poor quality, black points report on the projection in the latent space of all training examples, including the generated midpoints $\mathbf{z}_m$ used to train DDN. The latent space of the network based solely on MSE features two basins associated with "closed" and "open" states, separated by a region of poor quality models. In the network trained with both MSE and physics (DNN), the two states are connected by acceptable protein models.}
    \label{fig:landscape}
\end{figure*}

In addition to bonded terms (bonds and angles), protein conformations are also modulated by non-bonded interactions such as electrostatics and van der Waals. Including non-bonded potentials into the loss function along with the MSE and bonded terms further improved the quality of the transition state conformations (total errors 12.6\% $\pm$ 0.9, Figure \ref{fig:toterr}: Bonded + Non-bonded). 

To obtain more detailed information on the structural quality of each intermediate model, we assessed their Discrete Optimized Protein Energy (DOPE) score~\cite{shen2006statistical}. The DOPE score is an atomic distance-dependent statistical potential commonly used to assess experimentally determined and computationally predicted protein structures. Models generated by the neural network featuring non-bonded potentials in its loss function featured a slightly better DOPE score (-34173 $\pm$ 1648) than the network not featuring this term (-34104 $\pm$ 1703 at best), and the residue-level quality of the models generated by the best network was consistent with the profile of MurD crystal structure (PDB: 3UAG, see Figure S3 and Supplementary Material). 

To obtain a comprehensive view of the protein conformational space encoded in our neural networks, we calculated the DOPE score of protein models generated by regularly sampling the networks' latent space \ref{fig:landscape}. The latent space of the network trained solely with MSE featured two distinct regions, associated with MurD open and closed state, separated by a region of unacceptable quality (high DOPE score). In the network trained with our physics-based loss function, the whole conformational space appears near-convex, and all structures located between the "closed" and "open" basins had a low DOPE score. This was made possible by the addition in the training set of midpoints between training examples.

Overall, training our 1D CNN with a loss function featuring a combination of MSE, bonded and non-bonded terms yielded interpolations of good structural quality through conformations not represented within the training set.

\subsection{Neural network predicts a possible state transition path}

\begin{figure*}[ht!]
    \includegraphics[width=\textwidth]{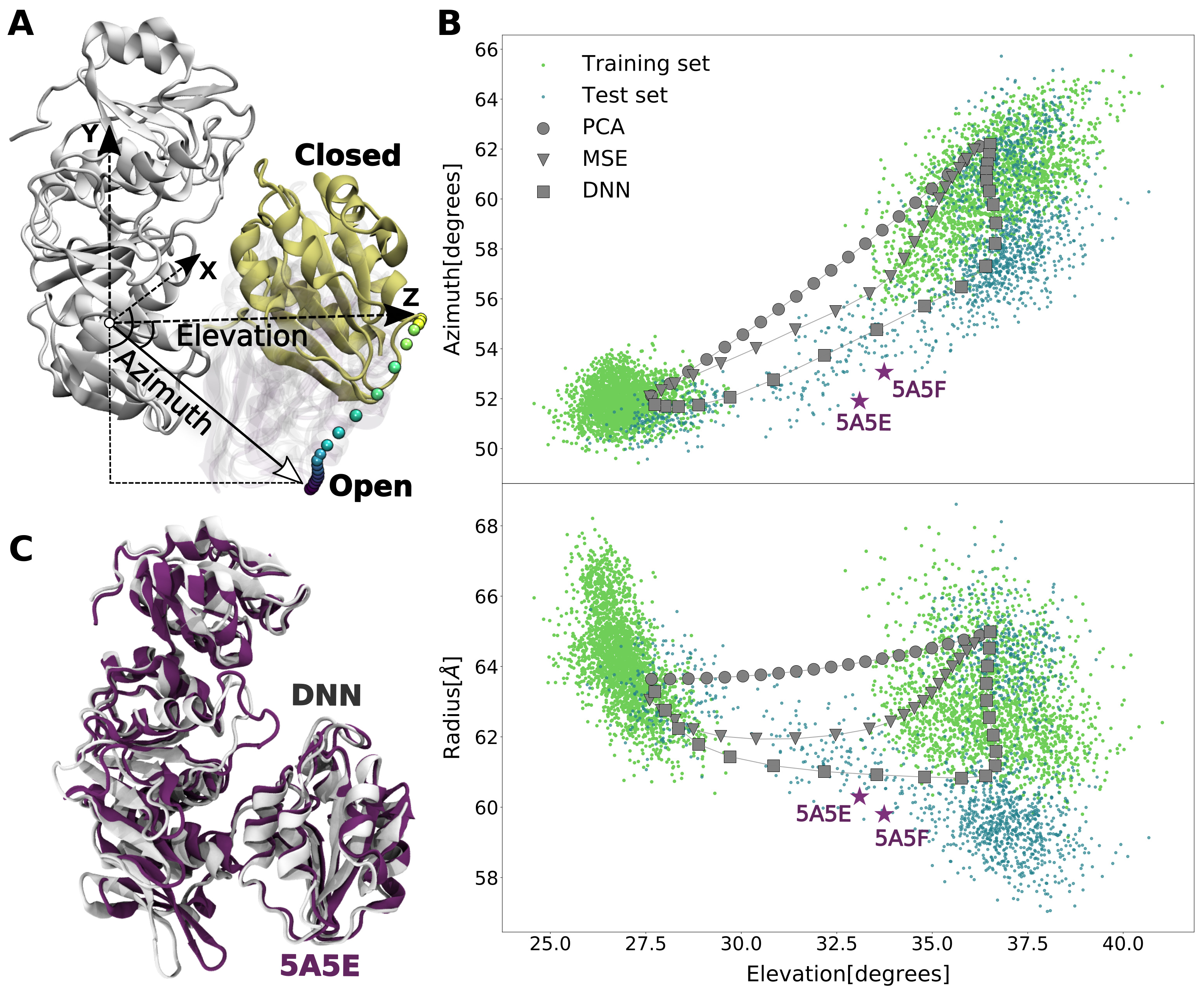}
    \vspace{-1em}\caption{State transition path prediction by the neural network. (A) Side view of MurD showing the transition of its mobile domain from the closed to open state (with the time evolution marked as beads coloured from yellow to violet). (B) The conformational change of MurD can be described in terms of spherical coordinates between its three domains (see Methods). We report the opening angle of each conformation in the training set (light green), test set (dark green), intermediate crystal structures (palatinate stars) as well as interpolations generated by PCA (gray circles), purely geometry-based neural network (MSE, grey triangles) and neural network combining geometry and physics (DNN, grey squares). The interpolation produced by DNN best reproduces the path described by the test set and transits in the vicinity of intermediate crystal structures. (C) Superimposition of intermediate MurD conformation 5A5E (in palatinate) and an intermediate conformation generated by the neural network (in white), with an RMSD of 1.2 {\AA}.}
    \label{fig:azim}
\end{figure*}

The switch of MurD between closed and open conformations involves the rigid-body rearrangement of one domain (residues 299 to 437) with respect to the rest of the protein (residues 1 to 298). We can readily characterize this movement by tracking the position of the centre of mass of this domain with respect of its connection to the rest of the protein and reporting it in spherical coordinates (Figure \ref{fig:azim}A). Describing the closed and open MurD MD simulations according to this metric reveals that the conformations of these two states are clearly distinct (two light green regions in Figure \ref{fig:azim}B). The MD simulation of MurD switching from closed-to-open state (hereon \textit{test set}) follows an irregular path: first, a concerted increase in elevation and azimuth opens the domain, leading to conformations closely resembling the crystal structure of the intermediates (RMSD of secondary structure elements equal to 1.16 and 1.12 {\AA} versus PDBs 5A5E and 5A5F, respectively), then an increase in azimuth and radius leads the domain to its final equilibrium position. Methods relying on purely geometric interpolations would be unable to predict such a two-step process. Indeed, the transition path between open and closed states generated by principal components analysis (PCA, as a linear interpolation within the simulations' eigenspace) traces a near-uniform far from what was observed in the test set. Considering more than 2 eigenvectors also had no significant improvement in the interpolations generated by PCA (Figure S5).

The transition path predicted by the neural network trained solely based on MSE is less uniform, but yet only poorly recapitulates the test set. The interpolation produced by the network trained considering MSE, bonded and non-bonded terms traces instead a path closely resembling the test set. The additional physics-based terms in the loss function not only helped in generating structures with correct bond lengths and angles, but also prevented an interpolation that would have required the protein domains to slightly compenetrate. Remarkably, one of the interpolated conformations had a backbone RMSD of 1.18 {\AA} from the intermediate crystal structure (PDB: 5A5E), a quantity comparable to that of the test set (Figure S6). 

Thus, our results show that in this application our 1D CNN trained with a combination of MSE and physics-based terms was capable of correctly identifying a possible non-linear conformational change between two distinct states.

\subsection{Transfer learning improves convergence}

\begin{figure*}[ht!]
    \includegraphics[width=\textwidth]{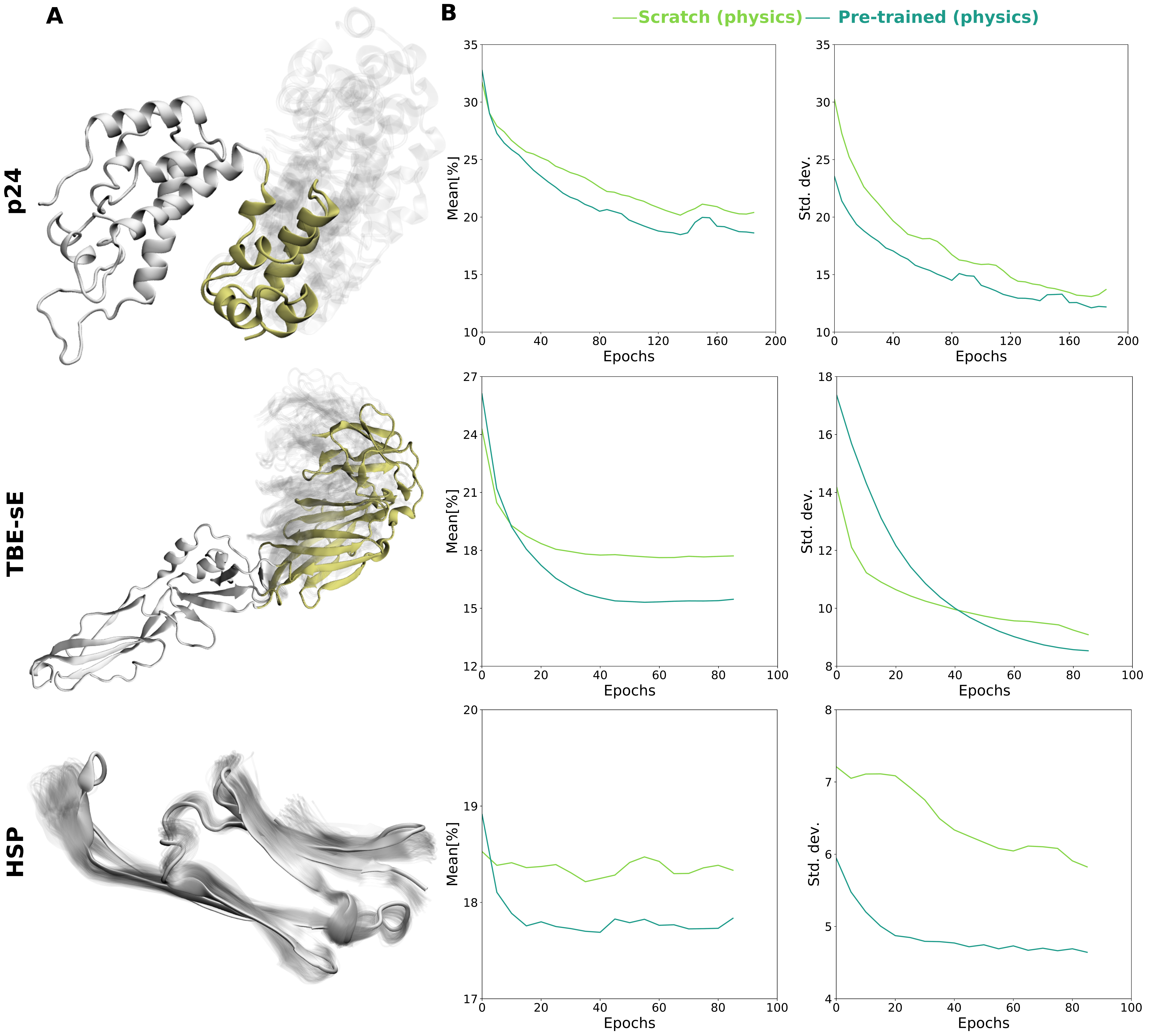}
    \vspace{-1em}\caption{Transfer learning of our best network trained on MurD to three different proteins. (A) Side view of p24, TBE-sE and HSP showing example alternative conformations with transparency and the highly mobile domains of p24 and TBE-sE in yellow. (B) Moving average (window of 3) mean and standard deviation of the total error percentage comparing the performance of the pre-trained network with its counterpart trained from scratch, both using the same loss function featuring both MSE and physics-based terms. The pre-trained network features a lower mean and standard deviation, as compared to the network trained from scratch. The slight increase in mean error at 150 epochs for p24 is caused by the triggering of non-bonded terms evaluation in the loss function (see Methods).}
    \label{fig:converge}
\end{figure*}

Training a neural network on a prohibitively small dataset can be possible if the network is pre-trained on a similar, larger dataset, a process called transfer learning. The convolutional nature of our network enables this training approach, as its architecture (and thus the number of its parameters) is independent of the number of atoms in the dataset under study. To assess the transferability of our trained network, we leveraged the MD simulations of three different proteins: the HIV-1 capsid protein (monomeric p24, PDB: 1E6J~\cite{monaco2000mutual}), the tick-borne encephalitis virus envelope protein E (TBE-sE, PDB: 1SVB~\cite{rey1995envelope}) and the small heat shock protein~$\alpha$B crystallin (HSP, PDB: 2WJ7~\cite{bagneris2009crystal}). These proteins differ from MurD in mass (24-kDa, 43-kDa and 10-kDa vs. 47-kDa, respectively) as well as fold and dynamics (Figure \ref{fig:converge}A). p24 consists of two rigid domains connected by a flexible linker, TBE-sE is highly elongated and features several long loop regions (e.g. residues 73 to 89 and 146 to 160), whereas HSP is small and features only local fluctuations.

From each simulation, we produced a dataset featuring 100 (and also 50 in the case of p24) representative structures (see Methods). For each dataset, we then trained our neural network in two different ways: from scratch, and by initializing its parameters with those of the best network trained on the MurD dataset using our physics-based loss function. Each neural network was assessed according to its capacity of interpolating between the two most different conformations (i.e. largest RMSD) in the training set.

In all the cases, the networks having transferred parameters converged faster than those trained from scratch, and generated interpolations with lower percentage errors both in terms of mean and standard deviation (Figure \ref{fig:converge}B). Furthermore, transfer learning yielded latent spaces featuring overall lower DOPE scores, including in regions extending slightly beyond the front defined by the training set (Figure S8). Results were particularly striking for p24, where we found no significant loss in the structural quality of intermediates generated by networks trained by transfer learning with either 50 or 100 structures (total percentage error of 15.0\% $\pm$ 3.5 vs. 16.3\% $\pm$ 4.2, respectively). Negative residue-level DOPE score indicated good quality atomic arrangements throughout the interpolation, with the only poor quality residues being located in correspondence of a long flexible loop (Figure S7, residues 75 to 87). p24 assembles into a circular homo-hexamer via a large-scale rearrangement of its two globular domains. While our training set featured a lowest RMSD of 4.0 {\AA} from the known bound state (PDB: 3MGE), the latent space could generate a slightly better model by local extrapolation (3.6 {\AA}, Figure S9).

Overall, these results demonstrate that our network can be trained with proteins of arbitrary size, and indicate that transfer learning enables faster training convergence, even when training data is limited, and leads to latent spaces better describing the conformational space of the protein.

\section{Discussion and conclusion}

Conformational dynamics are an intrinsic property of any protein, their magnitude depending on the required biological function. Several computational methods have been designed to sample the protein conformational landscape and predict existing states. Methods leveraging MD coupled with enhanced sampling methods, while helpful, are still limited by the timescales at which the transitions occur and the associated high demand for computational power. We have designed and trained a generative neural network with collections of protein conformations as a means to predict plausible intermediates between any of them. While here the network has been trained with structures from MD simulations, any source of structural information can in principle be used. Since as little as 50 structures were sufficient to produce physically correct models with our trained network, this opens the door to directly leveraging collections of experimentally determined atomic structures.

Generative networks \cite{kingma2013autoencoding,goodfellow2014gan,berthelot2018understanding} are effective when interpolating between existing data but, unless additional information is provided, their extrapolation capabilities are typically insufficient to generate plausible molecular structures. The additional difficulty when training a neural network with protein atomic coordinates is associated with the very high number of degrees of freedom to be handled (often >1000 individual $x$, $y$ and $z$ coordinates), making the training process arduous. To overcome these limitations, we have designed a 1D CNN architecture and introduced convolutional physics-based terms in the loss function. The convolutional architecture, already commonplace for image and text processing as well as bioinformatics tasks, is associated with a smaller number of trainable parameters and is independent of the number of degrees of freedom (and thus atoms) within the training set. Our physics-based terms in the loss function directly constrain learning of protein structures obeying essential force field potentials and, being fully convolutional, make the neural network faster to train (Figure S10). To further facilitate the training process, we have introduced a warping on the non-bonded potential terms (see Methods), leading to better gradients encouraging convergence in otherwise high Lipschitz constant conditions. To test our network we have purposely designed a set of hard tasks, by selecting the most different conformations from protein simulations featuring different ranges of conformational changes, and asking the neural network to generate suitable interpolations by leveraging only a $2$-dimensional latent space, as well as a limited number of training examples. 

Challenged with the flexible protein MurD, our network revealed a surprising non-linear transition path between two distinct states, closed and open. Remarkably, conformations along this path were both physically plausible and compatible with an existing MD simulation featuring a closed-to-open transition as well as with two crystal structures of MurD locked in intermediate conformations. While the MSE term in the loss function was key for the network to learn the global protein shape, the addition of physics-based terms was crucial to generating low-energy conformations, associated with a low DOPE score (Figure S3). The bonded terms (bonds and angles) helped fine-tuning local atomic arrangements, whereas the non-bonded ones (electrostatics~\cite{zhang2011role} and Lennard-Jones~\cite{adams2001bonding}) were a significant player in identifying a suitable transition path within the conformational space. The structures produced by our new network resulted to be both of lower energy and greater biological relevance than those obtained by PCA-based interpolations or by a neural network trained solely according to an MSE-based loss function (Figure \ref{fig:azim}B and C).
By analyzing the loss function values within the $2$-dimensional latent space of our physics-based neural network, we observed a clearly defined and near-convex minimum, despite the network having been trained with distinct conformational states (Figure \ref{fig:landscape}B).

An attractive feature of our convolutional network is that it can be trained with conformations of proteins of arbitrary amino acid sequence. This also enables transfer learning, whereby a pre-trained network is re-purposed to tackle a new though related task, expected to lead to improved generalisation and faster network optimization. In this context, we noted that different proteins still share common features (e.g. typical bonds length and angles, as well as the same sequence of atoms in the backbone), that should not be re-learned. We transferred the network trained on MurD to three different proteins (HIV-1 capsomer protein p24, tick-borne encephalitis virus envelope protein E and $\alpha$B crystallin small heat shock protein), for which we provided only a limited number of training examples (as low as 50). Remarkably, after few epochs the total error associated with structures generated by the pre-trained network dropped lower than that of structures produced by the network trained from scratch.

The intermediate structures generated by our neural network will feature overall low energy according to a subset of typical terms associated with molecular structures, namely their bond, angles, van der Waals and electrostatic interactions. Still, a more detailed analysis indicated that our neural network may generate suboptimal loop regions when these are highly flexible and the protein movement is dominated by larger domain-level conformational changes. Under those premises, the energy of predicted intermediates will not be an accurate estimation of energy barriers a conformational change is associated with. Thus, while capable of local extrapolation, the network should not be expected to predict new, completely unseen states. Nevertheless, knowledge of possible transition paths can provide guidance for the definition of appropriate reaction coordinates/collective variables in MD-based enhanced sampling schemes. Furthermore, as the neural network effectively transforms a discrete collection of protein conformations into a "conformational continuum", it can find applications in flexible protein-protein docking scenarios where, under the conformational selection paradigm, the ability of fine-tuning protein conformations to maximize their compatibility with a binding partner is desirable~\cite{degiacomi2019coupling}.

In summary, we have designed a new architecture with physics-based loss functions that significantly improve the synthesis of protein atomic structures in different conformational states. We have shown that non-linear biologically relevant transition paths can be predicted when synthesised interpolations must respect physical laws. The estimated non-linear transition path retains a suprising resemblence to the ground truth, and even more suprisingly this is achievable with only a small number of training examples. Further, our findings show that, through our fully-convolutional architecture, we can transfer features learnt from one protein to another, indicating the possibility of simultaneously training with conformations of multiple proteins, towards a network trained with the whole proteome.

\section{Methods}

\subsection{Network architecture}

Proteins are defined by their amino acid sequence, and each sequence maps onto an ensemble of possible three-dimensional atomic arrangements (conformations). The space of possible conformations associated with a specific sequence may be extremely reduced for a protein taking a single well-defined state, or broad for a flexible protein capable of interconverting between multiple states. As the proteome is vast and many proteins are resistant to most forms of experimental interrogation, only a relatively small collection of proteins have had at least one of their possible conformations revealed at atomic resolution. Furthermore, as the techniques characterizing molecular structures typically report on low energy conformations, transition states are undersampled proportionally to their associated energy barrier.
 
From a machine learning perspective, we can define the entire proteome as a distribution $p_d(\mathbf{x})$, where $\mathbf{x} \in \mathbb{R}^{3\times n}$ is a protein, and $m_d(\mathbf{x})$ a collection of conformations of a specific protein experimentally (e.g. nuclear magnetic resonance spectroscopy, x-ray crystallography) or computationally (e.g. Monte Carlo or MD simulations) determined. We wish to learn a low $m$-dimensional embedding $f : \mathbb{R}^{3\times n} \to \mathbb{R}^m$ that maps proteins onto the latent space, where sampling any point $\mathbf{z} \in \mathbb{R}^m$ and taking the inverse $f^{-1}$ yields a continuous space of physically plausible molecular structures. However, as the expected observations $\mathbf{x}\sim p_d(\mathbf{x})$ and $\mathbf{x}\sim m_d(\mathbf{x})$ relax into a subset of conformations~\cite{leach2001molecular}, the behaviour at the valley regions on the manifold (maxima in the energy landscape) is difficult to capture explicitly from the observations. 

Let $f(\mathbf{z}|\mathbf{x};\theta)$ be an encoder function with parameters $\theta$, and $d(\mathbf{\hat{x}}|\mathbf{z};\theta)$ be a decoder function that approximates the inverse $f^{-1}$ accordingly. The conventional approach is a simple reconstructive autoencoder that minimises the mean squared error loss $\mathcal{L}_{\textup{MSE}}$, where:
\begin{equation}
  \mathcal{L}_{\textup{MSE}}=\mathbb{E}_{\mathbf{x}\sim p_d(\mathbf{x})}\left[ \norm{d(f(\mathbf{x}))-\mathbf{x}}^2\right]
\end{equation}
This follows the geometry and probability distribution from which the dataset was collected and therefore fails to generalise, especially at undersampled regions associated with transition states. 

Proteins undergo conformational changes following lower energy paths in their energy landscape, where transition states are expected to be saddle points. The protein's expected energy, as determined by its atomic interactions, can be expressed as a loss function such as $\Psi(x,y) = \norm{x-y}^2$, defined as:
\begin{equation}
  \mathcal{L}_{\textup{phys}}=\mathbb{E}_{\mathbf{x}\sim p_d(\mathbf{x})}\left[ \Psi(d(f(\mathbf{x})),\mathcal{P}(\mathbf{x})) \right]
\end{equation}
where $\Psi$ is the error between the physical properties for the decoder (bond lengths, angles etc.) and the target properties $\mathcal{P}(\mathbf{x})$. However, as with the na\"ive autoencoder, this also fails to generalise at regions far from conformations provided as example. In principle, it is possible to enforce physics to be respected in regions outside the known conformational space, for example with a Gaussian prior in the latent encoding. However this makes an assumption on the distribution of the latent space.

Any midpoint along the geodesic (i.e. shortest path on the learned manifold) between any two protein conformations $(\mathbf{x_1},\mathbf{x_2})\sim m_d(\mathbf{x})$ will also be a protein of same connectivity and composition. Assuming a degree of convexity of the latent space, we can enforce our physics-based loss function at random midpoints between $\mathbf{x}_1$ and $\mathbf{x}_2$, sampled linearly from a uniform distribution:
\begin{equation}
  \mathcal{L}_{\textup{path}}=\mathbb{E}_{(\mathbf{x_1},\mathbf{x_2})\sim m_d(\mathbf{x}),\ t\sim U}\left[ \Psi(d(\mathbf{z}_m),\mathcal{P}(\mathbf{x}_1))  \right] 
\end{equation}
where $\mathbf{z}_m = (1-t)f(\mathbf{x}_1) + t f(\mathbf{x}_2)$ are midpoints in the latent space between two protein conformations. This enables physical characteristics to be enforced on the manifold between two points, but outside of the known sampled conformational space.

We note that, in principle, this approach allows for the training of a neural network with $m_d$ simulations of multiple proteins, whereby pairs (to interpolate between) are picked from the same simulation.

Putting this together, we can assemble the final loss as a weighted sum of the geometric term and the path term (which itself comprises of various individually weighted physics terms):
\begin{equation}
  \mathcal{L} = \alpha \mathcal{L}_\text{MSE} + \beta \mathcal{L}_\text{path}
\end{equation}

\subsection{Masked bonded loss}

Atoms in a protein structure file are listed residue-by-residue. So, the positions of atoms throughout the list are spatially coherent, where atoms involved in a common bond or angle are most likely adjacent in the list. We developed an efficient way to assess bonds and angles within a protein, by expressing $\Psi$ as differentiable 1D convolutions with a small number of pre-defined kernels that extract specific vectors relative for each atom in the list. The output of the convolution operator $\star$ is then multiplied by a binary mask $\mathbf{m}$ to filter only those vectors describing covalently interacting atoms.

For a given protein $\mathbf{x}$, the function $\mathcal{P}(\mathbf{x})$ returns sets of $k$ binary masks $\mathbf{m}^k$, target properties $\mathbf{t}^k$, and pre-defined convolution kernels $\mathbf{c}^k$. The error function $\Psi$ is defined as the sum of squared differences between the current properties $\rho$ and the target properties $\mathbf{t}$ multiplied by the mask, and normalised by the sum of the masked elements:
\begin{equation}
  \Psi(\mathbf{x},\mathbf{m}^k,\mathbf{t}^k, \mathbf{c}^k)=\frac{\sum (\mathbf{m}^k  \left(\rho(\mathbf{x},\mathbf{c}^k)  - \mathbf{t}^k \right)^2)}{\sum \mathbf{m}}
\end{equation}
where the final loss is the mean of each case $k$. 

To assess bond lengths, we convolve with a $2\times 1$ kernel for each (xyz) component:
\begin{equation}
  \rho_\text{dist}^2 = \sum_{j=1}^{3} \left( \begin{bmatrix}
    -1\\ 
    +1
    \end{bmatrix} \star \mathbf{x}_{j} \right)^2
\end{equation}
Similarly, for angles we take the dot product, using convolutions on the inner components:
\begin{equation}
  \rho_\text{angle} = \text{acos}\left( \frac{  \sum_{j=1}^{3}  \left( \mathbf{c}_1 \star  \mathbf{x}_{j} \right) \left( \mathbf{c}_2 \star \mathbf{x}_{j} \right) }{\norm{\mathbf{c}_1 \star \mathbf{x}_{j}}_2\norm{\mathbf{c}_2 \star \mathbf{x}_{j}}_2} \right)
\end{equation}
where the norms in the denominator are over the three components. For instance, to calculate the N-CA-C angle from atomic positions sorted as [N, CA, CB, C], we set $\mathbf{c}_1 = [0, -1, 0, 1]$ and $\mathbf{c}_2 = [1, -1, 0, 0]$. A list of all masks adopted in this work is provided in Supplementary Material (Table S1).

\subsection{Warped non-bonded loss}

The non-bonded potential $\rho_\text{nbp}$ consists of a sum of two terms, describing van der Waals and electrostatic interactions. $\rho_\text{nbp}$ is evaluated for any pair of atoms not involved in a mutual bond, angle or dihedral. Let $\mathbf{r}$ the distance between two atoms, we describe their van der Waals interaction as a 12-6 Lennard-Jones potential $\rho_\text{LJ}$:
\begin{equation}
  \rho_\text{LJ} = 4\epsilon \left[ \left( \frac{\sigma}{\mathbf{r}} \right)^{12}-\left( \frac{\sigma}{\mathbf{r}} \right)^{6}\right]
\end{equation}
where $\sigma$ is the equilibrium inter-atomic distance, $\epsilon$ the depth of the potential well. However, since side-chains atoms (except C$\beta$) are not used in the network training, only the repulsive term of the Lennard-Jones is applied. This is to avoid the protein structure to pack excessively, filling the voids left by missing side-chain atoms. In this work, we set $\epsilon = 1.0\textup{KJ}/\textup{mol}$ and adopted the combination rule $\sigma=(\textup{vdW}_\mathbf{a} + \text{vdW}_\mathbf{b})/2$ for the van der Waals radius to determine $\sigma$ for each atomic pair ($\mathbf{a}$ and $\mathbf{b}$). Thus, the Lennard-Jones potential is defined as: 
\begin{equation}
  \rho_\text{LJ} = 4\epsilon \left[ \left( \frac{\sigma}{\mathbf{r}} \right)^{12}\right]
\end{equation}

Electrostatic interactions are described by a Coulombic potential $\rho_\text{C}$:
\begin{equation}
\rho_\text{C}=\frac{\mathbf{q}}{\mathbf{r}}
\end{equation}
where $q$ is the multiplication of their respective charges. As $\lim_{\text{r}\to 0}\rho_\text{nbp}(r) = \infty$, the gradient descent becomes unstable at short inter-atomic distances. Clamping $\mathbf{r}$ causes the gradients to get stuck in the corners of the hypercube. Therefore, we approximate the non-bonded potentials by warping the input space $\mathbf{r}'\approx\mathbf{r}$ piecewise with an exponential. This is achieved equivalently and efficiently with the ELU~\cite{flevert2015fast} intrinsic activation function for some offset constant $k$, where:
\begin{equation}
  \mathbf{r}' = \text{elu}(\mathbf{r}-k,\alpha=1)+k
\end{equation}
such that the potentials now tend to high positive or negative values, without significantly altering the profile of the potential well (Figure S11).
We choose $k=1.9$ for $\rho_\text{LJ}$ and $k=0.4$ for $\rho_\text{C}$, giving large positive or negative y-intercepts for the expected upper and lower bounds for $\sigma$ and $\mathbf{q}$ accordingly.

\subsection{Implementation}
Atoms (points) of a molecular structure (such as in PDB file, the standard representation for molecular 3D structure data, Figure \ref{fig:conv}B) can be sorted in a list so that covalently connected atoms appear close to each other. 

The 1D CNN architecture was implemented in PyTorch with 12 layers (6 in the encoder and 6 in the decoder components). The first layer has 3 input channels (for each atom's 3D coordinates) and 32 output channels, where subsequent layers of depth $t$ have $\lfloor 32 \cdot {1.5}^{t} \rfloor$ input channels with batch normalisation and ReLU activations. Convolutions have size $4$ kernels with strides of $2$ and padding of $1$, halving the spatial dimension each layer. This means the molecule becomes like a progressively thicker but shorter `ribbon' whose activations correspond to more deep and abstract characteristics of the molecule, Figure \ref{fig:conv}C. The latent space was fixed to two dimensions using a 1D adaptive mean pooling layer to handle arbitrary length input sequences. 

We found that increasing the dimension of the latent space, adding an adversarial discriminating component, or adding residual layers led to only negligible generalisation improvements. The total number of parameters is 11,892,767 optimised using Adam with a learning rate of $1e^{-4}$ and weight decay of $1e^{-5}$ using a batch size of $5$ for 200 epochs and 1000 optimisation steps per epoch. Training the model takes $\sim$7 hours on an NVIDA TITAN Xp graphics card. In preliminary tests, we found that including non-bonded terms in the loss function only towards the end of the convergence of bonded terms (here, after 150 epochs) was beneficial to optimize the atomic arrangements.

Each neural network presented in this work was trained 10 independent times; all values reported (loss functions mean and standard deviation, as well as the DOPE score of resulting interpolated structures) are the average of these 10 repeats. The weight on the MSE penalty in the loss function was reduced to 0.1 from 0.5 for TBE-sE and HSP in view of their limited global dynamics. Their training runs converged by $\sim$100 epochs.

\subsection{Datasets}
MurD has been crystallized in its open (PDB: 1E0D~\cite{bertrand2000open}) and closed (PDB: 3UAG~\cite{bertrand1999determination}) states, as well as in intermediates between the two (PDB: 5A5E and 5A5F~\cite{vsink2016crystallographic} with a backbone RMSD of secondary structure elements equal to 1.12 {\AA} between them). The difference between these states comes from large scale conformational change of one of its three globular domains (residues 299 to 437) caused by substrate binding. Conformations from MD simulations of the closed and open states performed by some of the authors~\cite{degiacomi2019coupling} were used here as the training dataset (4420 conformations in total comprising of 2507 and 1913 from closed and open simulations, respectively). In order to evaluate the predictive performance of our network in interpolating between the diverse conformational states, a third set of MD simulations were performed with the ligand removed from the closed state (closed-apo)~\cite{degiacomi2019coupling}. We performed two additional repeats of this simulation using the same simulation protocol as~\cite{degiacomi2019coupling}. This produced a total of 1513 conformations representing a transition from the closed-to-open state unseen in either the closed or open state simulations. 

The simulations of p24 and TBE-sE were taken from~\cite{degiacomi2013macromolecular} and that of HSP from~\cite{degiacomi2019coupling}. We clustered all these simulations according to the RMSD of heavy atoms using an average-linkage hierarchical agglomerative clustering algorithm, then applying a cutoff criterion yielding exactly 100 centroids as representative structures. For p24, 50 of these representatives were randomly selected to test the performance of transfer learning on a smaller dataset.

For all datasets, we selected the backbone (C, C$\alpha$, N, O) as well as the side chain C$\beta$ atoms as representatives of protein structure. These are sufficient to describe the global conformation (fold) of a protein and the orientation of each side chain. The two extreme conformations in all cases (MurD, p24, TBE-sE and HSP) were selected by calculating an RMSD (backbone) matrix considering all the conformations of the training set and picking the pair with the highest RMSD. We generated 20 conformations interpolating between these extrema. The RMSD between these extreme conformations, an indicator of protein range of dynamics, was 10.2 {\AA} for MurD, 16.9 {\AA} for p24, 4.9 {\AA} for TBE-sE and 2.8 {\AA} for HSP.

\subsection{Percentage error calculation}
We used the equilibrium values of bonds and angles from the Amber ff14SB force field~\cite{maier2015ff14sb} to estimate the percentage error in the corresponding bonded parameters modelled by our network. The error over all the bonds and angles present in a conformation was summed to obtain the $\%_\text{TotalErr}$ as a measure of accuracy of the network in generating physically plausible protein structures: ${\%_\text{TotalErr} = \text{Err}_\text{bonds} + \text{Err}_\text{angles}}$. Similarly, the per-residue errors were calculated as the sum of errors associated with interactions involving any atom within a residue.

\subsection{Analysis of MurD opening angle}
The opening angles (azimuth and elevation) of MurD was calculated by aligning the protein along the stable domain (residues 1 to 298), centering the resulting alignment on the hinge between the mobile and stable domains (centre of mass of residues 230 to 298), and reporting the position of the centre of mass of the mobile domain (residues 299 to 437) in spherical coordinates. A schematic representation of the angles calculated is shown in Figure~\ref{fig:azim}A.

Visual inspections of the protein structures and related figures were done with VMD1.9.2~\cite{humphrey1996vmd} and PyMOL~\cite{PyMOL}. Graphs were plotted using matplotlib~\cite{Hunter2007pyplot}. 

\section{Acknowledgments}
The work was supported by the Engineering and Physical Sciences Research Council (EP/P016499/1).

\section{Author contributions}
V.K.R., C.G.W. and M.T.D. designed the research, developed the software, carried out the experiments, interpreted the results and wrote the manuscript.

\section{Declaration of interests}
The authors declare no competing interests.

\end{document}